# Spectrophotometric evidence for a metal-bearing, carbonaceous, and pristine interstellar comet 3I/ATLAS

Josep M. Trigo-Rodríguez[1,2]
Maria Gritsevich[3,4],
and Jürgen Blum[5]

[1]Institut de Ciències de l'Espai (ICE, CSIC), Campus UAB, Carrer de Can Magrans S/N, Cerdanyola del Vallès, 08193, Catalonia, Spain. E-mail: trigo@ice.csic.es
[2]Institut d'Estudis Espacials de Catalunya (IEEC), Esteve Terradas 1, Edifici RDIT, Of. 212, Parc Mediterrani de la Tecnologia (PMT),
Campus del Baix Llobregat – UPC, Castelldefels (Barcelona), 08860, Catalonia, Spain
[3]Faculty of Science, University of Helsinki, Gustaf Hallströmin katu 2, FI-00014 Helsinki, Finland
[4]Institute of Physics and Technology, Ural Federal University, Mira str. 19, 620002 Ekaterinburg
[5]Institut für Geophysik und extraterrestrische Physik, Technische Universität Braunschweig, Mendelssohnstr 3, D-38106 Braunschweig, Germany

**Abstract:** 3I/ATLAS is only the second confirmed cometary object known to enter the Solar System from interstellar space. Cosmogonic considerations suggest that this body may possess relatively high tensile strength and a substantial metal fraction. We present photometric observations along its inbound trajectory toward perihelion, together with a spectroscopic comparison to pristine carbonaceous chondrites from the NASA Antarctic collection. The spectral similarities indicate that 3I/ATLAS may be a primitive carbonaceous object, likely enriched in native metal and undergoing significant aqueous alteration during its approach to the Sun, experiencing cryovolcanism as we could expect for a pristine Trans-Neptunian Object. We propose that the combination of elevated metal abundance and abundant water ice can account for the unusual coma morphology and chemical products reported to date. To do so, corrosion of fine-grained metal grains can originate energetic Fischer-Tropsch reactions, generating specific chemical products into the coma that are not so common in other comets because most of them formed in the outer solar system and didn't inherited so much metal. Interstellar objects such as 3I/ATLAS provide rare opportunities to investigate physical and chemical processes in distant minor bodies of our own Solar System, including trans-Neptunian objects and Oort Cloud comets.

1. Introduction.

The mere existence of kilometer-sized natural bodies crossing interstellar space remained uncertain until the detection of 1I/'Oumuamua in 2017. In fact, the gravitational scattering of asteroidal and cometary bodies had been predicted long ago as a natural consequence of dynamical simulations of the turbulent epoch of planetary formation (see e.g. Morbidelli, 2008). In our own planetary system, the inward migration of Jupiter and Saturn played a particularly important role in reshaping the primordial small-body reservoir, producing significant gravitational scattering about 4 Ga ago (Morbidelli and Crida, 2007). A considerable fraction of the minor bodies formed in the protoplanetary disk were consequently ejected into interstellar space, becoming interstellar asteroids or comets.

The first clearly cometary interstellar object discovered was 2I/Borisov. Our own follow-up of this object with several instruments revealed a monotonic photometric trend without evidence of sudden brightness increases, or outbursts (Trigo-Rodríguez et al., 2024).

The third visitor from interstellar space, 3I/ATLAS (hereafter 3I), is even more intriguing. It was discovered on 2025 July 1 by the ATLAS survey using one of its Chilean telescopes (Tonry et al., 2018). 3I entered the Solar System with a remarkable trajectory and unusually high velocity, exemplifying the diversity of natural bodies capable of traversing interstellar distances. Understanding the nature of these visitors is crucial, particularly now that planetary defence has become a global scientific priority (Mainzer et al., 2021; Trigo-Rodríguez, 2022; Daly et al., 2023). Our current view of the main components of comets significantly improved after Stardust NASA mission, the only sample-return mission of cometary dust released from comet 81P/Wild 2 (Brownlee et al., 2006). In any case, interstellar visitors can challenge our models with bodies formed in other wide cosmochemical circumstances in distant planetary systems.

Dynamical analyses indicate that 3I is likely an ancient survivor, with an estimated age of 3–11 Ga (Taylor & Seligman, 2025). Pérez-Couto et al. (2025) further showed that 3I has not experienced close stellar flybys within the past 10 Myr, corresponding to about 500 pc, based on a careful study of stars included in the Gaia DR3 catalog. Such a long residence time in the cold regions of the Galaxy implies the formation of a near-surface mantle enriched in volatile ices and implanted dust grains. A fast-moving body in the interstellar medium (ISM) could have accumulated significant exogenous material on its surface, while at the same time experiencing continuous exposure to cosmic rays over billions of years.

From cosmogonic grounds, a body surviving for so long in the harsh ISM should have a significant mechanical strength. Observationally, 3I is also large: estimates place its diameter between 0.3 and 5.6 km (Jewitt et al., 2025), and its ~16 h rotation period is sufficient to distribute solar heating relatively uniformly across its surface (Santana-Ros et al., 2025). Its high incoming velocity suggests ejection through a close encounter in its parent planetary system, and although no stellar encounters are found within the last 500 pc, earlier close passages cannot be excluded. These considerations lead us to hypothesize that 3I/ATLAS may be a metal-bearing carbonaceous body. Several carbonaceous chondrite groups (CR, CH, CB), particularly those that are less aqueously altered, contain substantial FeNi metal. In the case of CH and CB chondrites, the high metal content may originate from a fragmental parent body formed by a giant collision between carbonaceous and iron-rich materials (Trigo-Rodríguez et al., 2025). The resulting structure would be a rocky-metal, thermally metamorphosed high-strength body covered by a porous regolith, somewhat analogous to asteroid (21) Lutetia (Moyano-Cambero et al., 2016), whose high bulk density, $(3.4 \pm 0.3) \times 10^3$ kg m$^{-3}$, implies a strong and metal-rich interior (Jin et al., 2018). On the other hand, Lutetia surface temperature measured by Rosetta was 245 K closer to our 3I effective temperature at 2.5 au (Coradini et al., 2011).

The evolutionary pathway of 3I must also be considered. Its long residence in the ISM likely produced a surface mantle enriched in volatile ices and organics, subsequently processed by intense cosmic-ray irradiation (Maggiolo et al., 2025).

Indeed, JWST/NIRSpec and SPHEREx observations reveal anomalous volatile ratios: $CO_2/H_2O = 7.6 \pm 0.3$, a $4.5\sigma$ deviation above values typical of Solar System comets, and $CO/H_2O = 1.65 \pm 0.09$. The unusually red spectral slopes may also be explained by prolonged exposure of the surface to energetic galactic cosmic rays.

Such implantation and cosmic-ray processing could partially obscure, at least in the first activation steps, the intrinsic composition of 3I. We propose that 3I may in fact be a metal-bearing transitional body probably similar to solar system Trans-Neptunian Objects (TNOs), specifically exhibiting a CR-like composition in which catalytic metal grains play a major role. In CR chondrites, FeNi metal is rapidly corroded by liquid water, producing clusters of magnetite observed in pristine aqueously altered specimens (Trigo-Rodríguez et al., 2019). The corrosion of metal can drive energetic Fischer–Tropsch reactions, potentially explaining the oxidized coma environment and the high CO and $CO_2$ abundances observed during 3I's approach to the Sun. We will explore this scenario in light of current observational evidence.

Since its discovery, 3I has continued to demonstrate a volatile-rich nature. As it approached the Sun, it developed a broad diffuse coma. Our photometric monitoring revealed a rapid brightness increase coincident with a heliocentric distance of 2.53 au, consistent with the onset of water-ice sublimation. This sudden brightening suggests that 3I contains abundant volatile ices, and entered a new sublimation regime due to solar heating.

We need to mention that ordinary comets are mantle-regulated. Most Solar System comets develop a dust mantle after repeated perihelion passages. This mantle insulates subsurface ice, throttles gas flow, self-regulates activity, and prevents runaway brightening (Brin and Mendis, 1979). Even if a comet experiences an outburst, the dust mantle often reseals rapidly, or the system equilibrates thermally, causing the brightness to return to a pre-outburst trend. 3I, being pristine and never thermally processed, likely has no such mantle.

As a consequence, if 3I had a so long journey through cold regions of our galaxy, we should expect having a thin ice-organics layer (Maggiolo et al., 2025), but preserving the pristinity of the object due to its extremely low thermal inertia and the insulating effect of the cover. A fast-moving body could experience significant amount of interstellar aggregate particles, and organics implanted on its surface (Maggiolo et al., 2025). In fact, these authors found from JWST/NIRSpec and SPHEREx that different ratios are anomalous: $CO2/H2O = 7.6+-0.3$ being a 4.5 sigma above solar system comets, and $CO/H2O = 1.65+-0.09$. In addition, the red spectral slopes might be explained by the long endurance interaction of 3I surface with energetic galactic cosmic rays.

Implantation of foreign materials, and their processing by cosmic rays during its long trip through our galaxy could have hidden the real nature of 3I. We will argue that current evidence suggests that 3I is a TNO-like primordial object, particularly exhibiting an affinity with a specific group of carbonaceous chondrites: the CR-chondrite group. Metal grains in this chondrite group can be quickly corroded by the action of water, transforming metal into clusters of magnetite that we observe in aqueously altered CR chondrites (Trigo-Rodríguez et al., 2019). In fact, the corrosion of these minerals originate energetic Fischer-Tropsch reactions that might also explain the oxidizing environment of the coma with CO and $CO_2$ gases first observed at early stages of 3I approach to the Sun. We will explore that scenario to explain current observational evidence.

From cosmogonic grounds, a comet surviving so long in the harsh interstellar medium (hereafter ISM), provides clues on the ability of bodies with relatively low or moderate strength to escape disruptive processes. We don't know its inner strength, but it must be large enough: 3I size was estimated to be between 0.3 to 5.6 km (Jewitt et al., 2025), while its rotation period of 16 h is fast enough to keep solar heat quite homogeneously (Santana-Ros et al., 2025). At the moment of writing these lines 3I has also survived its approach to the Sun. In fact, we should also take into account that its extreme interception velocity indicates it was probably launched from its planetary system through a close encounter. On the other hand, during its long journey we know it didn't encounter close stellar systems, during at least 500 pc, but we cannot rule out previous stellar encounters, even closer than it is now experiencing to the Sun in our solar system. Then, we think that 3I/ATLAS is a very pristine body that has a resemblance with metal-bearing CR chondrites. In any case, in meteorite collections there are some chondritic groups with a significant metal (and water) content without the need of invoking metal-rich asteroids. From less to more modal content in FeNi, we can find among the carbonaceous chondrites less aqueous altered groups like CR, CH, or CB chondrites. In the last two cases, the body could be fragmental, the by-product of a giant collision among a carbonaceous and an iron body. Large metal grains form the inner structure of the chondrite, enriching in a natural way its metal content (Trigo-Rodríguez et al., 2014, 2025). The result is a rocky-metal, thermally metamorphosed high-strength body, covered by a thick regolith rubble where metal grains can be abundant like e.g. asteroid (21) Lutetia (see e.g. Moyano-Cambero et al., 2016). The above mentioned asteroid has a high density and probably also high strength as the Rosetta ESA flyby measured its relatively high bulk density of $(3.4 \pm 0.3) \times 10^3$ kg/m$^3$, an important parameter for clues about the likely presence of metal grains in its composition and interior.

Since its discovery 3I has continued demonstrating its volatile rich content. During the approaching started to develop a diffuse coma, and during the pre-perihelion photometric follow-up presented here we noticed a fast increase in its magnitude of brightness. In fact, that sudden event occurred just at a heliocentric distance of 2.53 au, the one expected for significant sublimation of ices and close to the triple point temperature required for water sublimation. As we will describe properly in next sections, current scientific evidence demonstrates that 3I must contain significant amounts of water, and that the continuous increase in brightness points towards a new sublimation era motivated by the ice mantle in the entire body being activated and producing cryovolcanism.

The main goal of this paper is presenting a pre-perihelion photometric study of 3I/ATLAS, as a way to figure out the main volatile components of this comet. We will particularly focus in the increase in magnitude experienced at 2.5 au, in order to to introduce a basic physical scenario that reconciles its reflectance spectra with those of carbonaceous chondrites. This model is particularly important in order to gain insight on the main components expected for this fast visitor, gaining a reasonable explanation of its most likely nature to the general public. Despite media speculation, scientific observations can provide interesting clues if we use all our current knowledge about the nature of undifferentiated asteroids and comets.

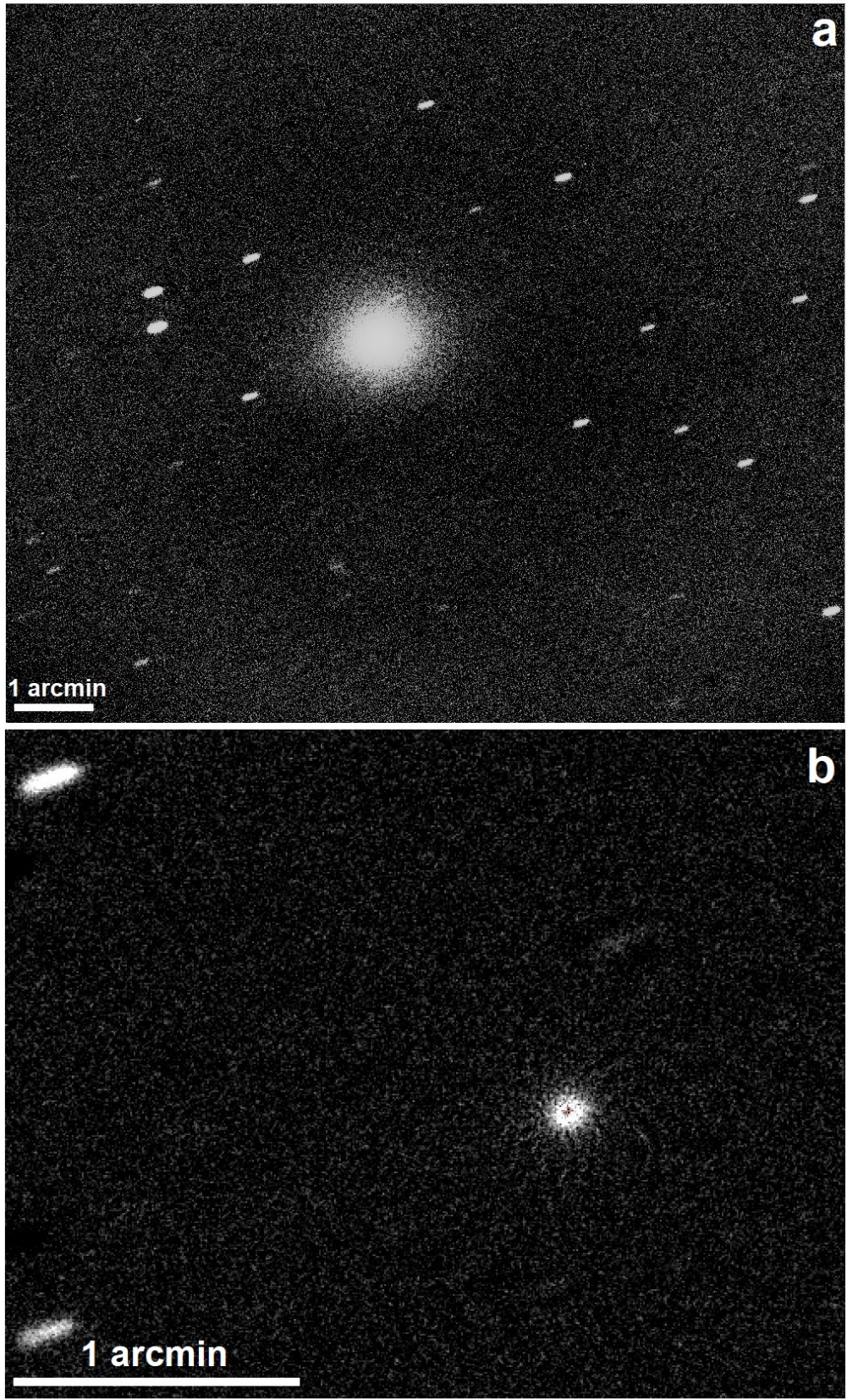

Figure 1 The concentrated, but diffuse appearance of 3I/ATLAS two weeks after its perihelion. a) A 6.5 minutes stacked image at 0.3 arcsec/pix. obtained from TJO telescope and MEIA2 camera on Nov. 13.216, 2025 (JD: 2460992.71), b) A 9º Larson-Sekanina filtered image rotated in the false nucleus (red cross) reveals several spiral-like jets surrounding the body. In both images, South is top, and West right.

2. Observations, and data comparison.

2.1. Photometric observations.

We observed comet 3I/ATLAS from several observatories in Catalonia, but mainly using *Telescopi Joan Oró (TJO)*, a 0.8 m robotic F/9.6 Ritchey-Chrétien telescope. It is operated remotely, but works in a Bortle 2-3 environment inside the *Observatori Astronòmic del Montsec (OAdM: www.oadm.cat)*. In addition, we also used a CDK 51 f/6.8 at Observatori de Pujalt, Barcelona. To study the photometric behavior of 3I/ATLAS we used these instruments, plus other smaller telescopes listed in Table 1.

TJO CCD detector is an iKon XL camera, with a back-illuminated 4k×4k chip manufactured by Andor. This CCD has a FoV of 27.5 x 27.5 arcmin$^2$, and provides a resolution of 0.4 arcsec for a pixel size of 15 μm.

Table 1. Observatories and instruments used in this study.

| **Observatory** | **MPC code** | **Telescope** |
| --- | --- | --- |
| Observatori de Pujalt (M04) | M04 | CDK 51 f/6.8 |
| TJO, Montsec (OdM) | C65 | RC 80 f/9.6 |
| Montseny Astronomical Obs. | B06 | SC 25 f/10 |

We used several filters with the CCD imaging. The photometry was performed using Astrometrica, being the results also tested with *LAIA (Laboratory for Astronomical Image Analysis)* a software successfully tested for obtaining high-precision stellar photometry (Juan-Samsó, et al., 2005).

The data reduction process consists of bias subtraction, flat-field correction and flux calibration using the IEEC Calibration and Analysis Tool (ICAT; Colomé et al. 2010). The time series photometry in the SDSS r band on the nights of 4, 6, 13, 16, 17, and 19 July was previously described in (Santana-Ros et al., 2025). For the rest of the imagery we used similar photometric reduction procedures than in the previous papers (Trigo-Rodríguez et al., 2008; 2010), and the photometry was measured for a circular aperture of 10 arcsec centered at the comet's false nucleus (Stetson, 2000). Using these standards, we were able to quantify the typical data accuracy to be better than 0.1 mag. and the results are plotted as a function of Julian Date and the heliocentric distance in Fig. 2.

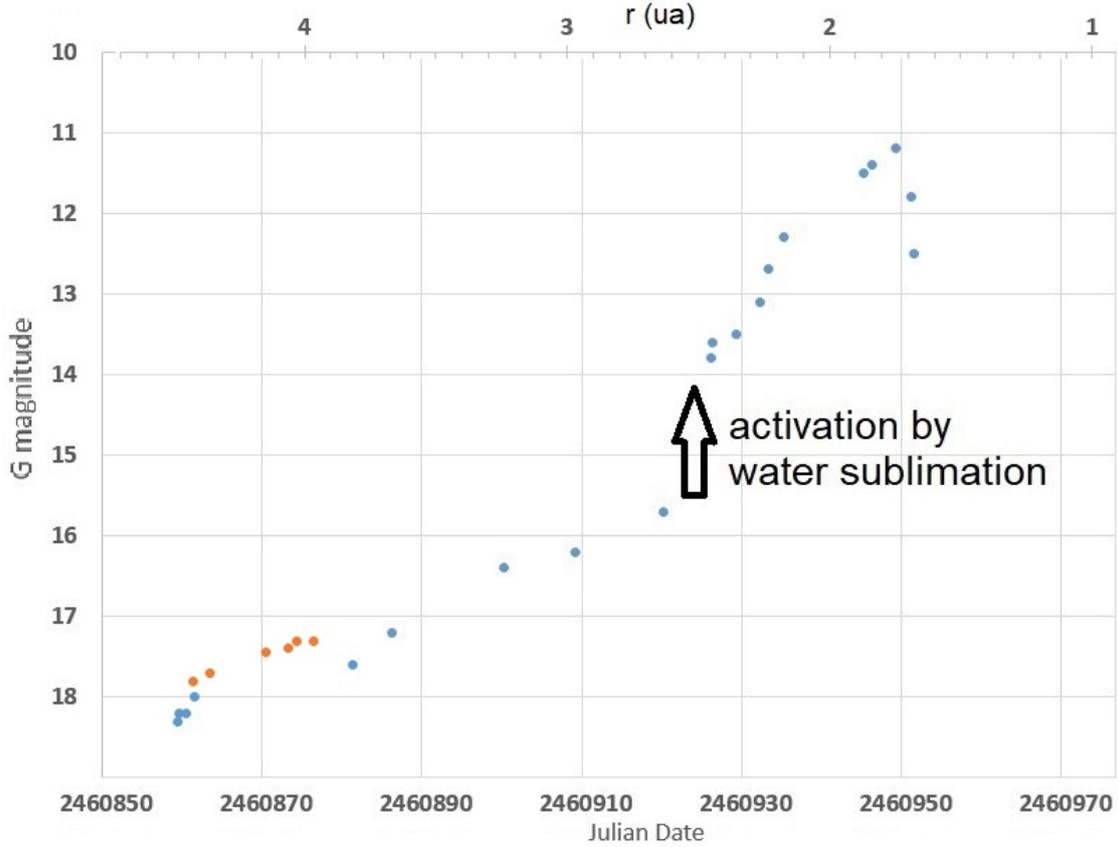

Figure 2. 3I g' magnitude as a function of the Julian Date (JD) and the heliocentric distance (in au) for preperihelion photometry. Error bars are not shown for clarity given that they are small. Data in orange are r' filtered.

2.2. 3I reflectance spectra, and comparison with meteorite spectra.

It is particularly relevant to compare the reflectance spectrum of 3I obtained at long heliocentric distances with those characteristic of pristine solar system objects. Fortunately, on the discovery date on July 1st was still a 4.55 au, and its activity was relatively minor, so first spectroscopic observations allowed obtaining 3I spectra that so far have been compared with comets and TNOs (Opitom et al., 2025; Yang et al., 2025). The previously presented photometric data suggests that at that heliocentric distance the sublimation of ices in 3I was not fully activated and probably the Sun's light on its surface was reflected as a function of the surface-forming minerals, with a relatively minor interference from dust in the coma.

In order to derive the main rock-forming minerals of 3I, we also consider relevant a comparison with meteorites, particularly with carbonaceous chondrites, given the link made previously by above mentioned authors with pristine TNOs and Centaurs (Opitom et al., 2025). For that reason, we performed a search among ICE-CSIC spectral database of CCs.

The 3I reflectance spectrum used here was obtained using the Gemini S GMOS instrument on 2025 July 5 (Yang et al., 2025). In that paper, the authors of the spectrum remarked that it exhibited a clear red slope, characteristic of D-type asteroids. That reflectance spectrum was kindly provided by B. Yang and we have normalized it at 550 nm to be compared with the reflectance spectra of carbonaceous chondrite (CC) meteorites, particularly from very well preserved CCs from the NASA Antarctic collection.

It is important to remark that for the spectral comparison; we have used several UV-NIR spectra obtained according to the technical procedures explained in Trigo-Rodríguez et al. (2012). We basically compared the reflectance spectra of different groups of carbonaceous chondrites with the 3I reflectance spectrum mentioned above. Carbonaceous chondrites are representative of pristine asteroids, and D, T and X spectral classes exhibit featureless spectra with characteristic red slopes (DeMeo et al., 2009). Those spectral classes are probably representing C-rich asteroids, or even comets.

The optical spectrum obtained by GMOS allows comparison in the VIS range (Fig. 3). We compared 3I's spectrum with the different chondrite groups, including extremely metal-rich groups, like CH and CB. No clear matches were found for most chondrite groups, and significantly differ because to get a similar slope it is usually required to increase the metal and other opaque mineral content to make them more reflective and red sloped, increasing their albedo (Trigo-Rodríguez et al., 2012). However, when comparing 3I with CR chondrites we found a reasonable similitude. In our spectral database we have three relatively pristine CR chondrites from the NASA Antarctic collection: EET 92159, GRA 95229, and LAP 02342. In addition, we also include a member of the Rumuruti group: PRE 95404. The reason for that choice was also the 10% albedo reported in the VIS range (Yang et al., 2025) that fits well the albedo of CR chondrites. In Figure 3 we plot the reflectance in percent, relative to barium sulphate used as reference (see Trigo-Rodríguez et al., 2012). It is important to remark that we plot the 3I spectrum normalized only with GRA95229 in order to keep the reflectivity information of the meteorites and exemplify that 3I behaviour is inside of the CR spectral diversity, in the studied window. We will discuss these results in Section 3.2.

Given the diversity of CR chondrite spectra, we decided to normalize 3I to GRA 95229 that exhibits an intermediate reflectance albedo of 10% at the given value of 550 nm. In such a way the spectra exemplify that 3I behaves well inside the limits of the CR reflectivity range. A quite similar 3I optical spectrum without distinguishable absorption bands and a similar red slope was obtained by Opitom et al. (2025). Their Fig. 3 is again a featureless spectrum with a remarkable red slope, in that case ending at 900 nm.

Despite of the obvious lack of connection among solar system parent body of CR chondrites and interstellar comet 3I, we find this similitude remarkable. It is obviously not perfect as the origin of the bodies is different, but it should be considered as a guide for the type of components we should consider in future modelling of 3I spectra. In next section, we will discuss what additional evidence we could infer from this resemblance with carbonaceous chondrites and its relevance to understand a bit better the nature, origin and composition of 3I.

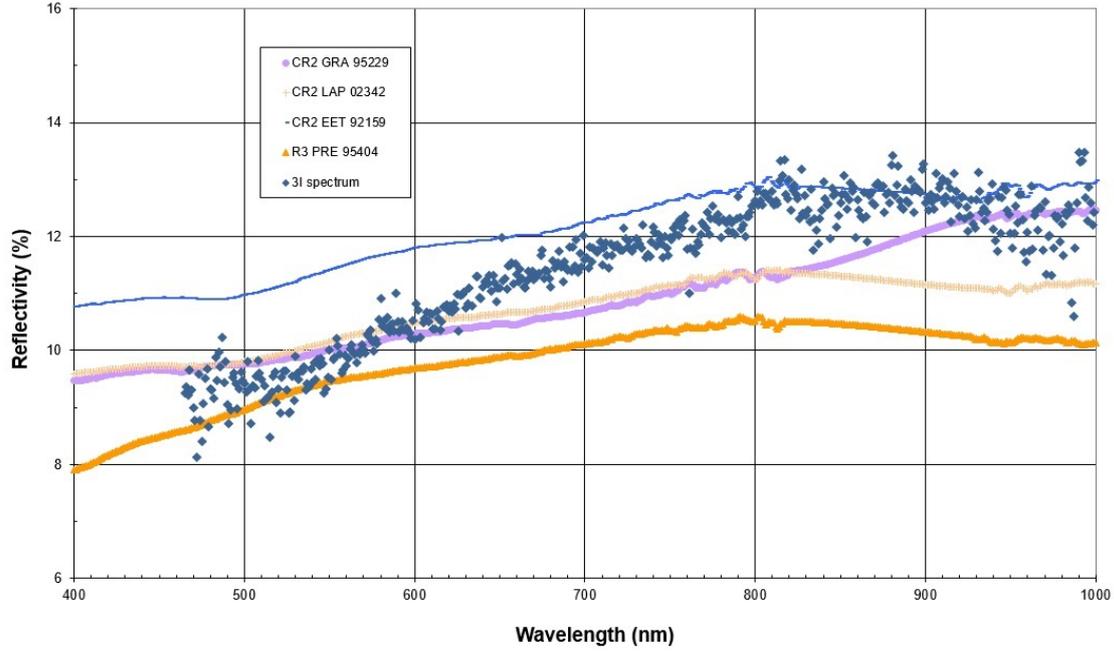

Figure 3. 3I spectrum (dots) and the reflectance spectra of three CR and one R chondrite discussed in the text.

### 3. Results and Discussion.

#### 3.1. Photometric study.

Concerning the observability period of 3I/ATLAS, it is clear that our manuscript contains photometric observations of the first months: July to November 2025. Afterwards the comet entered in a challenging geometry due to its vicinity to the Sun seen from our planet, so observations from the Earth were not possible. This merely casual circumstance was used by speculators, ufologists and other anti-science people to propose approaches that never considered the scientific evidence. 3I/ATLAS contains a significant amount of water. The evidence is in the photometry obtained, and other recent studies.

On the other hand, an additional demonstration that we are dealing with a cometary-like object is that we can fit 3I's photometric behaviour to a typical cometary magnitude curve, that might be fitted assuming a classical law in comet science:

$$m1 = m_0 + 5 \log (\Delta) + 2.5n \log (r) \qquad [1]$$

where $\Delta$ is the distance to Earth (in au) and r the heliocentric distance (in au). Given the light curve profile changes at 2.5 au, we should consider two different photometric laws. Before the onset of water ice sublimation [$m_0$= 12, n=2.2] and after the water activation [where $m_0$=7 and n=5.4].

As a better way to quantify the activity of dust in the coma, we used the measured photometric data and computed the Afρ values using a similar approach as in previous work (Trigo-Rodríguez et al., 2025). The Afρ computation was made from the formula relating the measured distances of the comet to the Sun and Earth (r, $\Delta$), the subtended size of 3I for our 10 arcsec aperture (ρ), and the Sun and comet magnitudes (Gillan, 2025, Page 16):

$$Af\rho = \frac{4r_h^2 \Delta^2}{\rho} \cdot 10^{0.4(m_\odot - m)} \qquad [2]$$

On the other hand, $m_\odot$ and $m$ are the apparent magnitudes of the Sun and comet respectively. $\rho$ is the photometric aperture, chosen here as 10 arcsec. For the absolute magnitude of the Sun we used the value of -26.97 for the R Cousins band given by Willmer (2018). On the other hand, to get the Afρ (corrected at zero phase angle) we used the Schleicher (2010) composite dust phase function, taking into account that the phase angle changed in the observational period in a wide range between 2.5° and 23º. Finally, the errors given in the Afρ values indicated in Table 2 were obtained using the method of partial derivatives, taking into account the uncertainties in the data given in equation [2].

As stated in Table 2, our photometric observations started with the comet discovery at a heliocentric distance of 2.448 au, and at a geocentric distance of 3.446 au. At that distance the selected aperture was subtending 25,000 km around the false nucleus, while at the end of the observational period the subtended region was about 18,000 km. Both are reasonable given the measured extent of the coma with larger instruments. On the first observed nights the measured Afρ values were in the range 280-290 cm (see Table 2). Over the following months the Afρ values increased progressively, particularly after the new sublimation era associated with water ice sublimation at 2.5 au. The specific Afρ values and derived propagation errors are listed in Table 2.

The most striking result from our pre-perihelion photometry is the sudden ~2-mag rapid activation phase surge observed at 2.5 au. This sustained increase is more consistent with a global activation event than with a classical cometary outburst, which typically returns to the original brightness level after the release of a localized pocket of volatiles (Gritsevich et al. 2025). In this case, the magnitude rise persisted, indicating a long-lived enhancement in activity rather than a transient outburst.

Such a brightness surge is compatible with the onset of vigorous water-ice sublimation, likely the dominant volatile species, as 3I crossed the heliocentric distance where water becomes thermally active. At these temperatures, near-subsurface water may also transiently become liquid in micro-environments within the porous matrix. For a body containing fine-grained metal, exposure to increasingly warm water would trigger significant corrosion reactions. These processes can release additional energy and gas, potentially amplifying the activity. Given the 16 h rotation period, the observed two-magnitude increase over a few days could represent an activation of the entire ice-rich surface, rather than a localized event.

To evaluate how unusual this behavior is for a comet, we computed the expected hemispheric-average and maximum surface temperatures of 3I/ATLAS following the approach described in Gundlach et al. (2011). First, we assume a solar flux of:

$$S(r) = 1361\,\text{W/m}^2 \left(\frac{1\,\text{au}}{r}\right)^2 \qquad [3]$$

where the heliocentric distance *r* is measured in astronomical units (au). Assuming a Bond albedo *A=0.1* for the surface material of 3I, the subsolar point on its surface absorbs:

$$P_{abs} = (1 - A)\, S(r) \qquad [4]$$

If we further assume that the heat conductivity of the surface material of 3I is low, the net heat flux into the subsurface is negligible against the re-emission of thermal radiation. The latter can be expressed by:

$$P_{em} = \sigma \cdot T_{eff}^4 \qquad [5]$$

With $\sigma$ and $T_{eff}$ being the Stefan-Boltzmann constant and the effective temperature of the sub-solar point, respectively. Consequently, using [4] the latter equation can be written as follows:

$$T_{eff} = \left(\frac{(1-A)S(r)}{\sigma}\right)^{1/4} \qquad [6]$$

The effective temperature for r = 2.53 au, results in $T_{eff}=241$ K. For a slow rotator and/or low thermal inertia, we can also calculate the hemispherical average temperature by equating the absorbed flux over the full hemisphere with its emitted flux and gets:

$$T_{eff} = \left(\frac{(1-A)S(r)}{2\sigma}\right)^{1/4} = 202\ \text{K}$$.

If these temperatures are also representative for the water ice close to the surface of 3I, they result in a peak (sub-solar point) and hemispherical average fluxes of 0.034 kg/s·m² and 0.00027 kg/s·m², respectively, provided that the gas permeability of the dust matrix is high. Consequently, both temperatures are apparently too low for liquid water. The triple point of water is at 273 K and 600 Pa pressure. We can neither reach 273 K, nor the 600 Pa pressure here at 2.5 au, so perhaps the sublimation of a more volatile ice was involved. In any case, even when liquid water cannot be efficiently sublimated at that heliocentric distance, perhaps it can be in the close subsurface as a consequence of the thermal wave producing a thin interstitial layer driving the alteration processes with reactive minerals.

Afρ values range from about 250 cm during the week of the discovery to values five times larger when the comet crossed the heliocentric distance of 2.5 au. Concerning the dust production rates, our data are consistent with the r−2 variation expected of a strongly volatile material found by Jewitt and Luu (2025). These authors also found a preponderance of 100 μm grains in the coma of 3I, and they emphasized that the interstellar visitor is characterized by exhibiting a heliocentric index of activity (n=1.9) typically asteroidal.

On the other hand, the significant increase at 2.5 au can be also explained if that distance marks the onset of energetic reactions. Particularly, if large particles are energetically released into the coma a significant outgassing power is needed, and we will discuss in the next section the main drivers, particularly invoking the activation of cryovolcanism.

Table 2. Observational data: Julian date, R magnitude, heliocentric distance (r), geocentric distance (Δ), and Afρ (cm).

| Date | Julian Date | g' magn. | r magn. | r (au) | Δ (au) | Afρ (cm) |
|---|---|---|---|---|---|---|
| 2025-07-02.979 | 2460859.48 | 18.3 | - | 4.448 | 3.446 | 231±21 |
| 2025-07-03.229 | 2460859.73 | 18.2 | - | 4.441 | 3.422 | 227±18 |
| 2025-07-04.020 | 2460860.52 | 18.2 | - | 4.415 | 3.416 | 218±16 |
| 2025-07-05.063 | 2460861.55 | 18 | - | 4.382 | 3.387 | 203±17 |
| 2025-07-05.063 | 2460861.56 | - | 17.8 | 4.380 | 3.387 | 205±18 |
| 2025-07-06.938 | 2460863.44 | - | 17.7 | 4.316 | 3.333 | 208±20 |
| 2025-07-13.946 | 2460870.45 | - | 17.5 | 4.085 | 3.160 | 213±18 |
| 2025-07-16.910 | 2460873.40 | - | 17.4 | 3.986 | 3.084 | 216±15 |
| 2025-07-17.910 | 2460874.40 |  | 17.3 | 3.953 | 3.061 | 227±29 |
| 2025-07-19.916 | 2460876.42 |  | 17.3 | 3.888 | 3.019 | 225±26 |
| 2025-07-24.813 | 2460881.31 | 17.6 | - | 3.726 | 2.928 | 315±21 |
| 2025-07-29.750 | 2460886.25 | 17.2 | - | 3.565 | 2.845 | 334±27 |
| 2025-08-12.750 | 2460900.25 | 16.4 | - | 3.114 | 2.679 | 342±32 |
| 2025-08-21.760 | 2460909.26 | 16.2 | - | 2.832 | 2.617 | 240±21 |
| 2025-09-01.781 | 2460920.28 | 15.7 | - | 2.492 | 2.574 | 368±28 |
| 2025-09-07.750 | 2460926.25 | 13.8 | - | 2.318 | 2.557 | 1580±145 |
| 2025-09-07.781 | 2460926.28 | 13.6 | - | 2.317 | 2.557 | 1560±133 |
| 2025-09-10.771 | 2460929.27 | 13.5 | - | 2.230 | 2.551 | 1375±121 |
| 2025-09-13.781 | 2460932.28 | 13.1 | - | 2.145 | 2.544 | 1342±119 |
| 2025-09-14.813 | 2460933.31 | 12.7 | - | 2.117 | 2.542 | 1247±108 |
| 2025-09-16.781 | 2460935.28 | 12.3 | - | 2.063 | 2.538 | 1216±103 |
| 2025-09-26.760 | 2460945.26 | 11.5 | - | 1.808 | 2.514 | 921±83 |
| 2025-09-27.760 | 2460946.26 | 11.4 | - | 1.783 | 2.511 | 887±77 |
| 2025-09-30.750 | 2460949.25 | 11.2 | - | 1.715 | 2.501 | 865±74 |
| 2025-10-02.760 | 2460951.26 | 11.8 | - | 1.672 | 2.493 | 847±71 |
| 2025-10-02.760 | 2460951.49 | 12.5 | - | 1.666 | 2.492 | 829±68 |

### 3.2. Spectral comparison

On cosmogonical grounds, the apparent similarity between 3I and Solar System CR chondrites is plausible and may help us understand the change in behaviour once the object crossed the heliocentric distance where water ice becomes thermally unstable. The CR chondrite group likely accreted significantly later than other carbonaceous chondrite groups, and several studies show that the fine-grained matrix in CRs is compositionally distinct and not simply the byproduct of chondrule-forming processes (Schrader et al., 2014; Davidson & Schrader, 2023). We cannot rule out that the CR parent body incorporated a substantial amount of water, accreted in the form of hydrated minerals (Trigo-Rodríguez et al., 2022).

This raises the possibility that 3I is an ancient, largely undifferentiated body that retains primitive outer-disk materials. While we cannot confirm this directly, the hypothesis is consistent with its spectral properties and observed activity. The CR chondrites used for comparison in Fig. 3 contain significant native metal, sulphides, and unusually rich inventories of complex organic compounds. Notably, the CR2 meteorite

GRA 95229 contains an order of magnitude more amino acids than the CM2 Murchison meteorite, likely generated through Fischer–Tropsch–type reactions catalysed by metal grains on the parent body.

CR chondrites also contain xenoliths — small inclusions of foreign material derived from other distant bodies. One of the meteorites used in Fig. 3, LaPaz Icefield 02342, recovered by ANSMET in 2002, contains a cometary or transneptunian xenolith with extremely unusual mineralogy and abundant presolar grains and insoluble organic matter (Nittler et al., 2019). This evidence indicates that the CR parent body formed in the outer Solar System, in a region where it interacted with early Centaur- or TNO-like material. Since Centaurs are fragments of TNOs scattered inward by the giant planets, both populations preserve outer Solar System materials that were also exposed to interstellar implantation. Most TNOs exhibit NIR absorption bands associated with water ice, while the visible spectrum is neutral and featureless. Such combination is usually explained adding different mixtures of organics, amorphous carbon and water ice (Fornasier et al., 2004). Remarkably, LAP 02342 provides a close spectral match in the visible region to 3I, particularly between 700 and 1,000 nm, perhaps indicating that this meteorite fragment is among the least thermally or aqueously altered samples in the CR chondrite population, perhaps representing a good proxy for an ancient TNO.

The continuous and increasing outgassing exhibited by 3I strongly suggests that its outer layers, possibly containing implanted ices and organics, host significant volatile reservoirs. Its activity differs from that of typical Solar System comets, particularly in showing an unusually oxidizing coma. Early observations reported $CO_2$ and CO as dominant species. Later, Salazar-Manzano et al. (2025) detected the onset of CN emission in mid-August, followed by OH detection at 2.90 au (Xing et al., 2025). Such an early release of water implies a high $H_2O$ production rate of ~40 kg s$^{-1}$. Consequently, 3I is among the few comets with confirmed OH emission beyond 2.5 au—distances where water-ice sublimation from the nucleus is typically inefficient. This behaviour is consistent with long-term implantation of ices and organics in the outermost layers. Those layers may have been processed by cosmic rays into a carbon-rich amorphous crust, which could explain why 3I appeared spectrally carbon-chain depleted (Salazar-Manzano et al., 2025). As 3I approached the Sun, the progressive heating likely fractured this irradiated crust, revealing a more porous and friable aggregate structure beneath that released ices and dust more efficiently. Sublimation, combined with exothermic reactions between volatiles and embedded reactive minerals, would then drive the ejection of hydrated material into the coma. These fragile aggregates quickly disintegrate under rising temperatures, contributing secondary water to the coma.

High-resolution imaging of the false nucleus after perihelion shows intense activity with multiple jets extended until at least 50,000 km from the nucleus, some displaying wavy or filamentary structure (Fig. 4). Metal grains and sulphides within carbonaceous chondrites are known to be extremely reactive in the presence of moderately warm water (Guo et al., 2007), but water vapour can also condense in the interstitial regions where metal grains are embedded. Just as example, static aqueous alteration is common in carbonaceous chondrites and usually exemplifies that water was heterogeneously distributed, producing localized aqueous alteration in water-containing carbonaceous asteroids (Zolensky and McSween, 1988; Trigo-Rodríguez et al., 2019). In CR chondrites, for example, water played a key role in mobilizing certain reactive elements like Ni and S, initially present in metal and sulphide grains contained into the fine-grained matrix, or in the interior of chondrules as exemplified by the aqueous alteration mineral products identified by Trigo-Rodríguez et al. (2013).

At 2.5 au, water sublimation likely permeated the subsurface. Previous work has shown that the thickness of the diffuse surface layer of water ice increases considerably at around 210 K, which leads to an enhanced mobility of the water molecules on the surface (Gärtner et al. 2017). This finding is further corroborated by the fact that the collisional behaviour of microscopic water-ice particles changes drastically from below to above ~200 K (Gundlach & Blum 2015). Both results indicate that liquid water may be present in small quantities at temperatures well below the triple point and achievable at 2.5 au. Whether these are sufficient to aid exothermic chemical reactions, still needs to be seen. If under such conditions, energetic Fischer–Tropsch-type reactions between water and reactive metal grains have contributed substantially to the energy budget of 3I, they may have enhanced both gas release and dust entrainment from its fine-grained interior due to cryovolcanism, a process occurring in solar system TNOs (Geissler, 2015; Guilbert-Lepoutre et al., 2020; Fagents et al., 2022). Cryovolcanism has been also naturally invoked to explain the photometric changes experienced by comet 174P/Echeclus (Wesołowski, 2020).

In any case, 3I continued its approach to the Sun. On the perihelion on Oct. 29$^{th}$ it was at a minimum heliocentric distance of 1.356 au, so for the subsolar point, we obtained T=329 K (clearly above the triple point of water), and for the whole hemisphere T=277 K also above the required temperature to melt water. These temperatures are close to the values computed for the aqueous alteration of pristine carbonaceous chondrites (Guo et al., 2007).

Then, we have learned from it that the onset of activity from an icy-rich and C-rich primordial body as 3I that initially is in a state of dormancy for long time, is expected to be quite fast after the sublimation of the most volatile ices. Our data suggest that it probably happened at ~2.5 a.u. forming liquids that quickly evolved to the subsurface, entering in contact with the silicate–ice and metal-ice interface, then producing localized thermal activity. As we state for the expected products, the formation and composition of the fluid reservoir has implications for the composition of the resulting cryomagmas, and outgassed components (Fagents et al., 2022).

Concerning the interaction between such oxidizing liquids and the reactive metal, and sulphide components, previous studies have demonstrated that these reactive minerals contained in carbonaceous chondrites have catalytic properties and are able to reproduce complex organics in very short timescales, particularly in presence of hot water and formamide (Rotelli et al., 2016). The relevant action of Fischer-Tropsch reactions also explains the gas products identified in further studies (Cabedo et al., 2021).

Interestingly, one of the most remarkable and unique features in the coma of 3I/ATLAS is the presence of Ni, with prominent absorption lines (Hutsemékers et al., 2025). If we invoke Fischer-Tropsch ongoing in the interior of 3I we can get a possible explanation for the overabundance of Ni in the coma. One example of the reactions we consider exemplified by a recent study of the interaction of schreibersite ($Fe_2NiP$) under the presence of water (Pantaleone et al., 2022). Their simulations indicate that water, when molecularly adsorbed, binds preferentially through an O−metal bond to Ni and less so to Fe. Consequently, we think that we are observing a bias introduced by the formation of volatile species that are probably launched by the intense outgassing of the body, producing localized cryovulcanism. That mechanism produces the jets, and releases Ni-rich dusty products into the coma (Fig. 4).

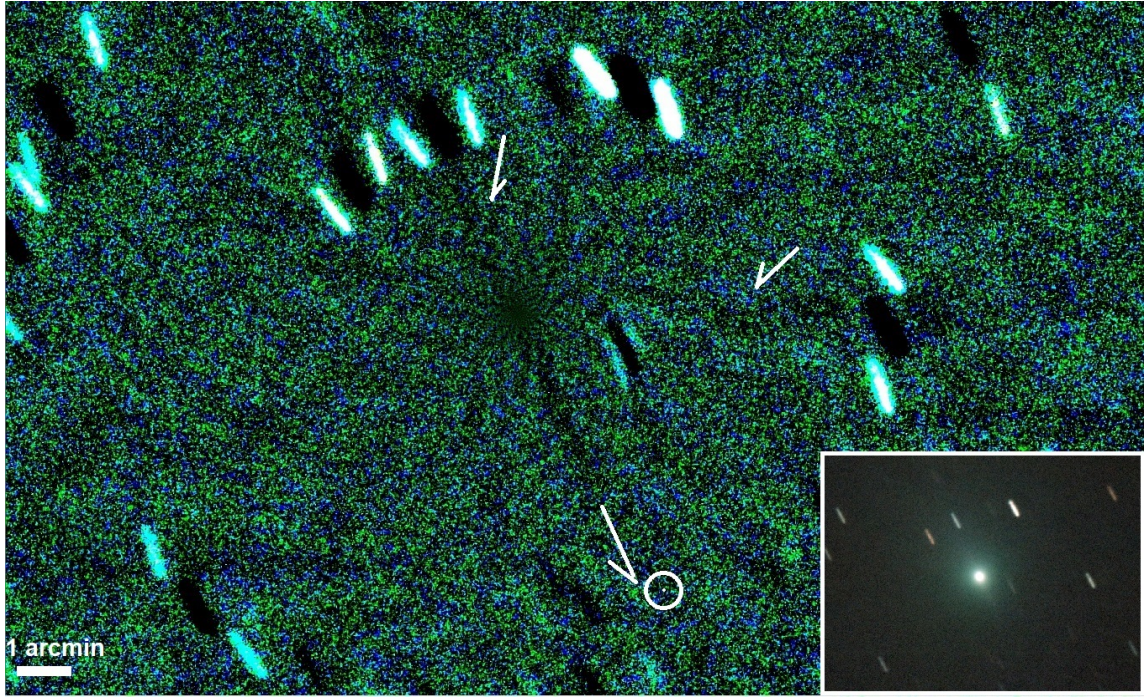

Fig. 4. Nov. 19th image obtained by Pau Montplet from Breda (Girona) using a C6 telescope at f:7. Original image is in the lower right inset, while a false colour 0.6º wide Larson-Sekanina filtered image at 9º shows 3I in negative to remark the antitail pointing to the subsolar point, and several jets getting out from the false nucleus. Two additional arrows mark wavy structures in the jets. Image resolution is 0.7 arcsec/pix.

If we are right about the ubiquity of Fischer-Tropsch reactions in 3I, we could also predict having some phosphorized species like $PO_4^{3-}$, so we could find its specific emission bands in the MIR, for example around its extended emission band around 9.5 microns. In addition, a porous undifferentiated body experiencing Fischer-Tropsch processes could release unusual species for comets formed in more reducing environments, like e.g. the $CO_2$ and $H_2O$. If we are correct, we predict that coma of 3I might contain more alcohols than hydrocarbons, making evident the bias caused by the preferred participation of Ni over Fe due to Fischer-Tropsch reactions. Spectral studies during the next few weeks after perihelion should take into account the main radicals produced by the degradation of 3I minerals by highly oxidizing species.

Our preferred explanation about the nature of 3I is invoking a transitional C-rich TNO-like body that was launched from its ancient planetary system during a close encounter with a giant planet, or with its star. We expect that a 1 km sized transitional body like that could have a rough mass of $6\times10^{11}$ kg, and according to our study of the chemical abundances of CCs, in particular the CR group contains a 25.7 wt% of Fe. On the other hand, concerning Ni, we obtained about 7,500 ppm, or writing this in weight per cent we obtain 0.75 wt% (Trigo-Rodríguez et al., 2025). These numbers exemplify that a km-sized CR-like asteroid might contain $4.5\times10^9$ kg of Ni, most of them as microns-sized grains (Fig. 5a). If 3I left its planetary system at an earlier stage, it might be younger in the sense of less collisionally processed than the CR parent asteroid. Consequently, the bulk porosity could be much higher than solar system proxies, as we expect from formation models and laboratory simulations (Blum et al., 200, 2014;

Trigo-Rodríguez and Blum, 2009). In the circumstances experienced by 3I in approaching the Sun covered by a mantle of ice, we envision that its porous nature can facilitate water soaking the body and generating catalytic reactions when entering in contact with pristine subsurface regions of the asteroid. Such processes could also release volatiles and produce metal-rich dust as observed for 3I.

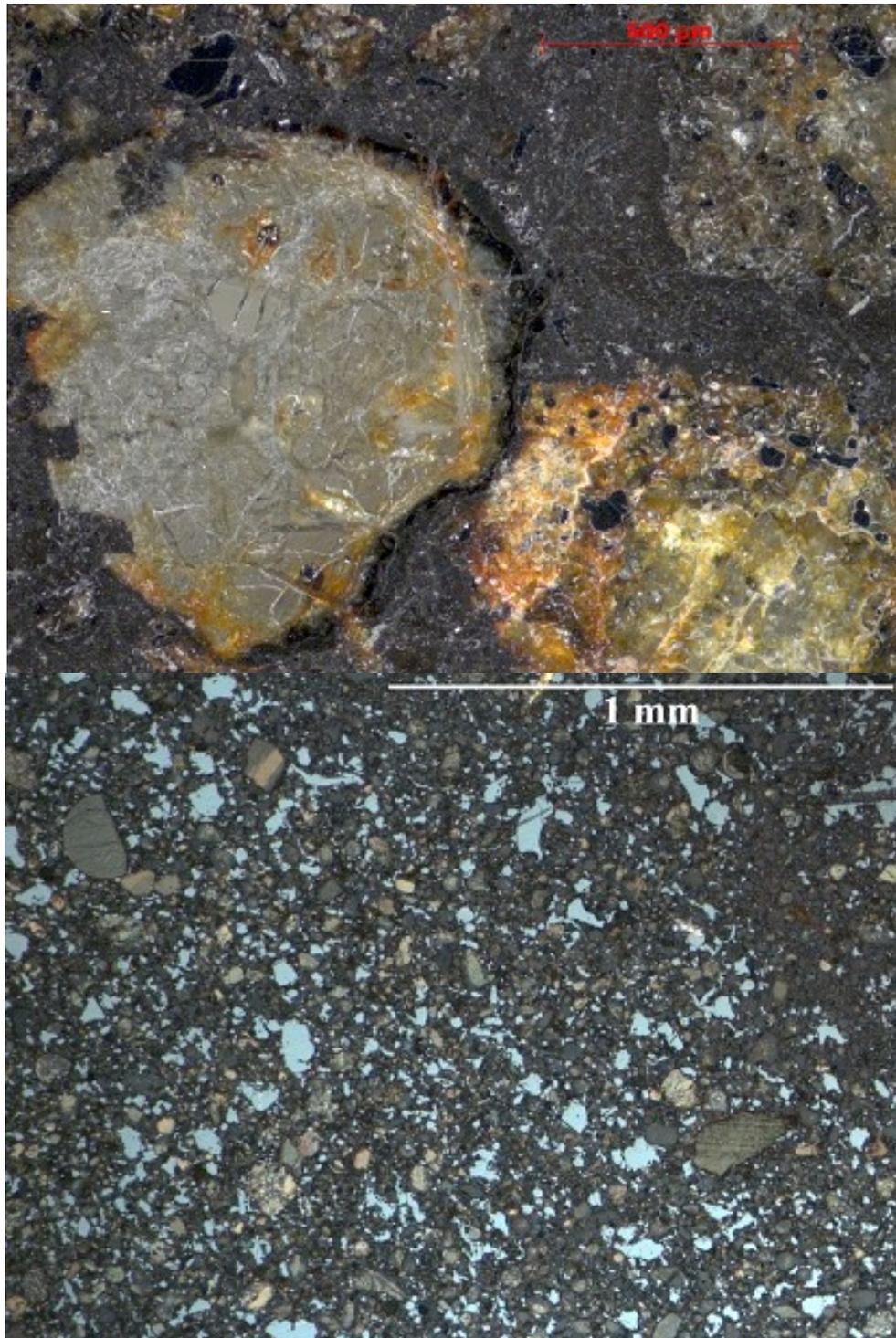

Figure 5. Reflectance images using a Zeiss Axio Scope of two thin sections of the carbonaceous chondrites whose spectra was compared with 3I. a) CR2 chondrite GRA 95229 where olivine rich chondrules are set into a dark, opaque-rich, matrix. and b) Reflectance image of PCA91467 where the overabundant metal grains can be seen in light blue color. Chondrules are rarely complete, so it has a fragmental structure.

It is important to consider that the presence of extremely reactive minerals could also explain the evolutionary behaviour observed in Centaurs. Being fragments of distant objects, once they reach heliocentric distances in which their ices are activated might experience significant catalytic reactions in their interiors if they contain fine-grained metal (obviously exposition of fresh volatile ices could promote similar behaviour). Some Centaurs are known to experience bright cometary outbursts like e.g. 29P/Schwassmann-Wachmann 1 (Trigo-Rodríguez et al., 2008, 2010). Such energetic processes will remove their surfaces, then becoming much less red. The existence of a colour bimodality in the Centaur population is known that probably reflects the bodies that have experienced significant loss of surface materials. Then, active Centaurs exhibit a less red color index (see e.g. Kokotanekova et al. 2025). In fact, Opitom et al. (2025) suggested that the red slope of 3I is probably the consequence of its long stay in the interstellar medium, where its surface was strongly affected by energetic cosmic ray radiation. Then, 3I exhibits similitudes with non-active Centaurs and long period comets, objects probably experiencing a similar irradiative processing by cosmic rays in the distant regions that they visited during their aphelia, close to the heliopause.

Concerning the observed dust production (Afρ) of 3I, the typical values are also similar to the typical for Jupiter-family comets (JFCs), a dynamical subclass of short-period comets in low-inclination orbits having their probable origin in the Kuiper Belt region (Gillan et al., 2024). If we compare 3I photometric evolution with the typical for JFCs according with that study, the Afρ maximum values could be about 100 days after perihelion. Our inner coma observations (Fig. 1 and 4) reveal that the interstellar visitor remains significantly activated.

We are not tempted to think that some of the observed features, particularly the dominant Ni in the coma, could be better explained by a rocky-iron or iron differentiated asteroid as core of 3I. The expected strength and homogeneity expected for such asteroids made us rule out that scenario. On the other hand, if we consider 3I as a mixture of carbonaceous transitional asteroid, plus metallic bodies, we could also find some examples in our own solar system. In such a case, we should refer to CH chondrites, sharing similitudes with carbonaceous chondrites, but containing more metal in mass probably due to be a product of a catastrophic collision with a metal-rich projectile (see Fig. 5b and 6) (Moyano et al., 2016). Theoretical studies propose that comet nuclei can be also processed by impacts during the early stages of the evolution of the protoplanetary disk, but it is unclear that such effects should be noticeable nowadays (Weissman et al., 2020). Then, it is not impossible that radial mixing was able to produce impacts of metal-rich asteroids with porous objects, ending as CH-like materials. Unfortunately, to our knowledge, the only relatively well preserved CH chondrite is PCA 91467, so our capacity to study the reflective diversity for this chondrite group is limited.

Although is not our favorite explanation, in the case in which 3I was the product of a collision, and ended containing big chunks of iron, then we should consider that particularly some iron meteorite groups IIC, IID, and IIAB in our solar system contain a really reactive mineral called schreibersite ($Fe_2NiP$). In a similar way than native metal, schreibersite could react really fast under the presence of water as mentioned before. Then additional catalytic reactions of this mineral could reinforce the corrosion of the body, providing additional chemical energy for developing an unusually rich metal-rich coma.

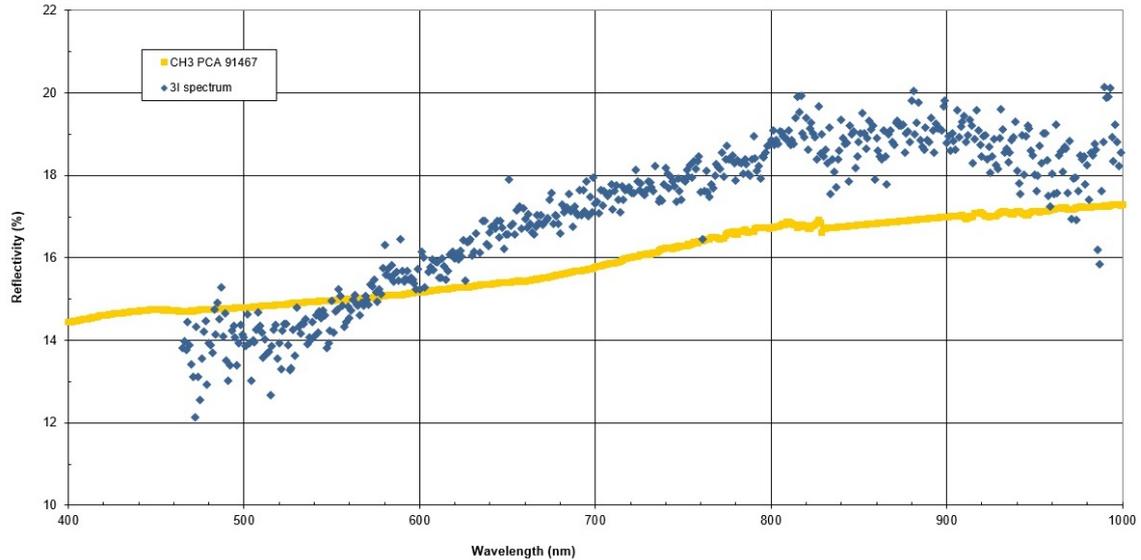

Figure 6. 3I spectrum (dots) and the reflectance spectra of CH chondrite PCA 91467 (yellow).

4. **Conclusions.**

Interstellar object 3I/ATLAS provides a unique opportunity to investigate the composition, activity, and alteration processes of minor bodies formed around other stars. Our photometric and spectroscopic study suggests that 3I is a primitive, metal-rich carbonaceous body undergoing aqueous alteration, linking its behavior to the chemical diversity seen among distant Solar System small bodies. During its pre-perihelion approach, 3I exhibited activity patterns that reflect its internal composition and volatile inventory. Cometary activity strongly influences coma morphology, producing jets and curtains of dust and gas. Our analysis leads to the following conclusions:

a) Our photometric follow-up demonstrates that 3I entered a new sublimation regime at 2.53 au, consistent with the onset of the sublimation of the ice mixture forming this pristine body.

b) The corresponding ~2-mag brightness increase at 2.5 au, followed by the rapid development of a diffuse coma, confirms the activation of near-surface volatile components, even when water-ice sublimation probably was not fully achieved, except perhaps in the near subsurface as consequence of reaching a more localized pressure and temperature. Cryovolcanism is known to occur on pristine ice-rich and C-rich bodies in the outer solar system, and 3I behaviour has an overall similitude to what we should expect for a pristine TNO during a close approach to the Sun.

c) We infer an average near-surface temperature at 2.5 au of 202 K, yielding peak (subsolar) and hemispherical-average water-vapor fluxes of 0.034 kg s$^{-1}$ m$^{-2}$ and 0.00027 kg s$^{-1}$ m$^{-2}$, respectively, assuming a permeable dust matrix. Solar heating, together with exothermic reactions involving reactive minerals, likely powered the observed coma development. A volatile ice mixture should be envisioned to drive those reactions.

d) Reflectance spectroscopy of carbonaceous chondrites provides important clues to the fine-grained materials that may compose 3I. Its featureless spectrum and positive slope was previously proposed resembles those of some TNOs and Centaurs, and now we demonstrate that it also lies among the reflectance spectra of CRs, among the most pristine carbonaceous chondrites, probably associated with these distant minor bodies.

e) Despite its unknown extrasolar formation environment, 3I shows its closest spectral affinity with CR and CH chondrites, both of which exhibit featureless red spectra—likely due to the presence of metal grains, sulphides, and other opaque phases. It means that early stages of planetary formation could produce similar types of materials, even in really remote locations of our galaxy.

f) These comparisons support the presence of significant fine-grained metal in the interior of 3I. Together with the anomalous activation at 2.5 au and the reported Ni enrichment in the coma—unusual for typical comets—this suggests that 3I is a transitional object containing substantial amount of metal, and volatile species, at difference of usual solar system comets.

g) Our metal-bearing scenario envisions a porous carbonaceous body with moderate water content. In this framework, water vapour interacting with newly exposed metal grains may drive catalytic reactions that release additional energy and contribute to the observed Ni outgassing.

h) The inbound trajectory and observing geometry of 3I highlight important lessons for planetary defence. Its steep inclination (175°) and unfavourable solar conjunction made ground-based follow-up extremely challenging, emphasizing the need for multi-platform observational capabilities.

i) The ground-based ongoing international Asteroid Warning Network (IAWN) observational campaign can be extremely useful to learn about the photometric behaviour of interstellar comets. In any case, space-based telescopes, particularly working in the IR, are needed to increase our capabilities of detecting interstellar comets.

j) Dedicated missions such as *ESA's Comet Interceptor* should therefore be considered a high priority for the direct exploration—and ultimately the sampling—of future interstellar visitors.

j) Sampling-return missions from interstellar comets seem challenging and perhaps only reliable to collect grains as made by Stardust NASA. In fact, our post-perihelion observations show increasing activity on 3I's surface, producing energetic jets where water penetrates fractured rock and corrodes pristine metal-bearing regions. The images clearly indicate that the entire nucleus is activated and outgassing from multiple sources, with especially strong curtains of material near the subsolar point.

Interstellar visitors like 3I/ATLAS continue to challenge and refine our understanding of planetary-system formation and the chemical evolution of small bodies. Each newly discovered object reveals unexpected properties that test and expand current models. Future intercept missions will be essential for visiting, and directly sampling these rare messengers and unlocking the record they carry from distant planetary systems.

Acknowledgements

We acknowledge support from the Spanish Ministry of Science and Innovation, research project PID2021-128062NB-I00 (PI: JMTR), funded by FEDER/Ministerio de Ciencia e Innovación – Agencia Estatal de Investigación.. The programme of development within Priority-2030 is acknowledged for supporting the research at UrFU. We thank Herbert Raab for support with *Astrometrica* software, and Pau Montplet i Sanz for providing 3I observations. Some of the TJO images at OAdM were scheduled by Toni Santana (UB) who kindly provided the r' photometric measurements, others by the JMTR. The Joan Oró Telescope (TJO) at the Montsec Observatory (OdM) is owned by the Catalan Government and operated by the IEEC. We also thank photometric data by T. Lehmann & M. Jager shared in the comets-ml.